\documentclass[twocolumn,showpacs,preprintnumbers,amsmath,amssymb,prb,aps]{revtex4-2}

\usepackage{graphicx}
\usepackage{bm}
\begin{document}
\title{Anomalous Nernst effect in the noncollinear antiferromagnet Mn$_5$Si$_3$}
\author{Christoph S\"{u}rgers$^{1}$}
\email{christoph.suergers@kit.edu}
\author{Gerda Fischer$^{1}$}
\author{Warlley H. Campos$^{2}$}
\author{Anna Birk Hellenes$^{2}$}
\author{Libor \v{S}mejkal$^{2,3}$}
\author{Jairo Sinova$^{2}$}
\author{Michael Merz$^4$}
\author{Thomas Wolf$^4$}
\author{Wolfgang Wernsdorfer$^{1,4}$}
\affiliation{$^1$Karlsruhe Institute of Technology, Physikalisches Institut, P.O. Box 6980, D-76049 Karlsruhe, Germany}
\affiliation{$^2$Institut für Physik, Johannes Gutenberg Universität Mainz, D-55099 Mainz, Germany}
\affiliation{$^3$Institute of Physics, Czech Academy of Sciences, Cukrovarnick\'{a} 10, 162 00 Praha 6, Czech Republic}
\affiliation{$^4$Karlsruhe Institute of Technology, Institute for Quantum Materials and Technologies, P.O. Box 3640, D-76021  Karlsruhe, Germany}
 
\begin{abstract}
Investigating the off-diagonal components of the conductivity and thermoelectric tensor of materials hosting complex antiferromagnetic structures has become a viable method to reveal the effects of topology and chirality on the electronic transport in these systems.
In this respect, Mn$_5$Si$_3$ is an interesting metallic compound that exhibits several antiferromagnetic phases below 100 K with different collinear and noncollinear arrangements of Mn magnetic moments. 
Previous investigations have shown that the transitions between the various phases give rise to large changes of the anomalous Hall effect. 
Here, we report measurements of the anomalous Nernst effect of Mn$_5$Si$_3$ single crystals.
Below 25 K we observe a sign change of the zero-field Nernst signal with a concomitant decrease of the Hall signal and a gradual reduction of the remanent magnetization which we attribute to a subtle rearrangement of the magnetic moment configuration at low temperatures.
\end{abstract}
\date{\today}

\maketitle

\section{Introduction}
Antiferromagnetic materials have attained a renaissance in condensed-matter research due to technical advantages like low stray-fields and ultrafast switching compared to ferromagnets and leading to the development of a new field coined antiferromagnetic spintronics \cite{jungwirth_antiferromagnetic_2016,baltz_antiferromagnetic_2018}. 
A particular class of materials are antiferromagnets in which the magnetic moments of atoms are ordered in a noncollinear fashion. They often exhibit a nonzero Berry phase curvature leading to an emergent electromagnetic response, which can be harnessed for practical purposes \cite{nakatsuji_large_2015,nagaosa_anomalous_2010,machida_time-reversal_2010,chen_anomalous_2014,kubler_non-collinear_2014}. 
In noncollinear antiferromagnets and spin liquids a nonzero Berry phase curvature gives rise to an unusually large anomalous Hall effect (AHE). 
Like in ferromagnets, the intrinsic part of the AHE is obtained by integration of the Berry phase curvature of occupied electronic bands over the entire Brillouin zone \cite{xiao_berry-phase_2006,xiao_berry_2010}. 
Similar to the AHE, its thermoelectric counterpart, the anomalous Nernst effect (ANE), generates a voltage transverse to the heat flow and magnetization. 
This transverse thermopower provides a measure of the Berry phase curvature only at the Fermi energy $E_{\rm F}$ \cite{xiao_berry-phase_2006,xiao_berry_2010}. 
The AHE and ANE have been the subject of intense research and development in recent years and hold great promise for practical applications in the fields of spintronics and thermoelectronics. 
Both effects can be extraordinary large in chiral antiferromagnets like Mn$_3$Sn and Mn$_3$Ge despite their tiny magnetization \cite{nakatsuji_large_2015,kiyohara_giant_2016,nayak_large_2016,ikhlas_large_2017,wuttke_berry_2019,chen_anomalous_2021}. 
Here, the enhanced Berry-phase curvature is associated with the existence of Weyl points near the Fermi level where the Berry curvature diverges. 

The intermetallic compound Mn$_5$Si$_3$ is a noncollinear antiferromagnet which has gained attention due to unusual thermodynamic and electronic transport phenomena \cite{gottschilch_study_2012,surgers_large_2014,surgers_anomalous_2016,surgers_switching_2017,biniskos_spin_2018,das_observation_2019}. 
Mn$_5$Si$_3$ has hexagonal crystal structure (space group $P6_3/mcm$) with two inequivalent Mn lattice sites Mn$_1$ and Mn$_2$ at room temperature and undergoes two structural phase transitions toward orthorhombic symmetry below 100 K.  
\begin{figure}
	\includegraphics[width=\columnwidth]{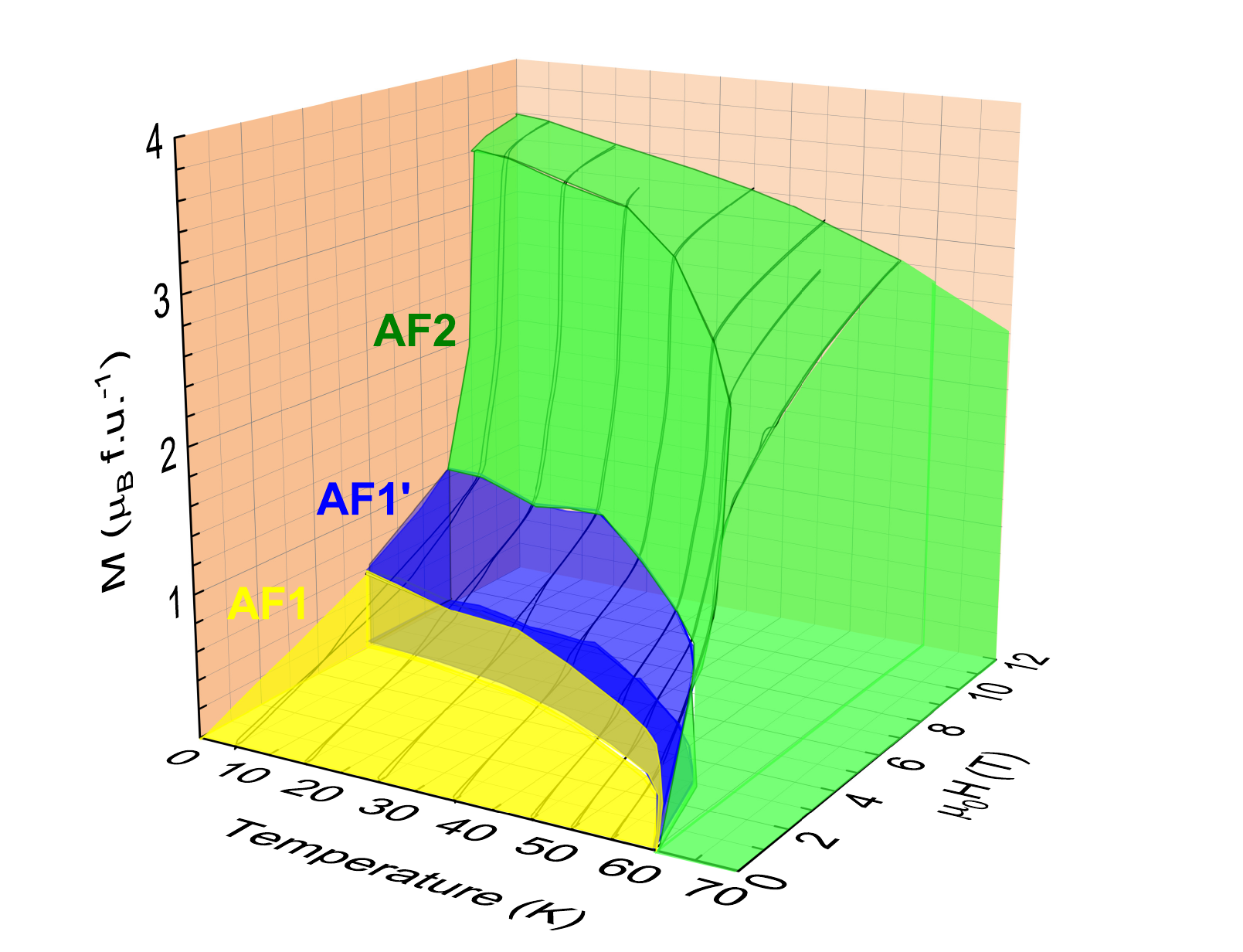}
	\caption[]{Magnetic phase diagram and magnetic moment of Mn$_5$Si$_3$ obtained from individual magnetization curves, see Fig. \ref{fig2}(a). AF1 and AF1’ exhibit noncollinear antiferromagnetic structures while AF2 is a collinear antiferromagnetic phase.}
	\label{fig1}
\end{figure}

\begin{figure}
	\centering
	\includegraphics[width=\linewidth]{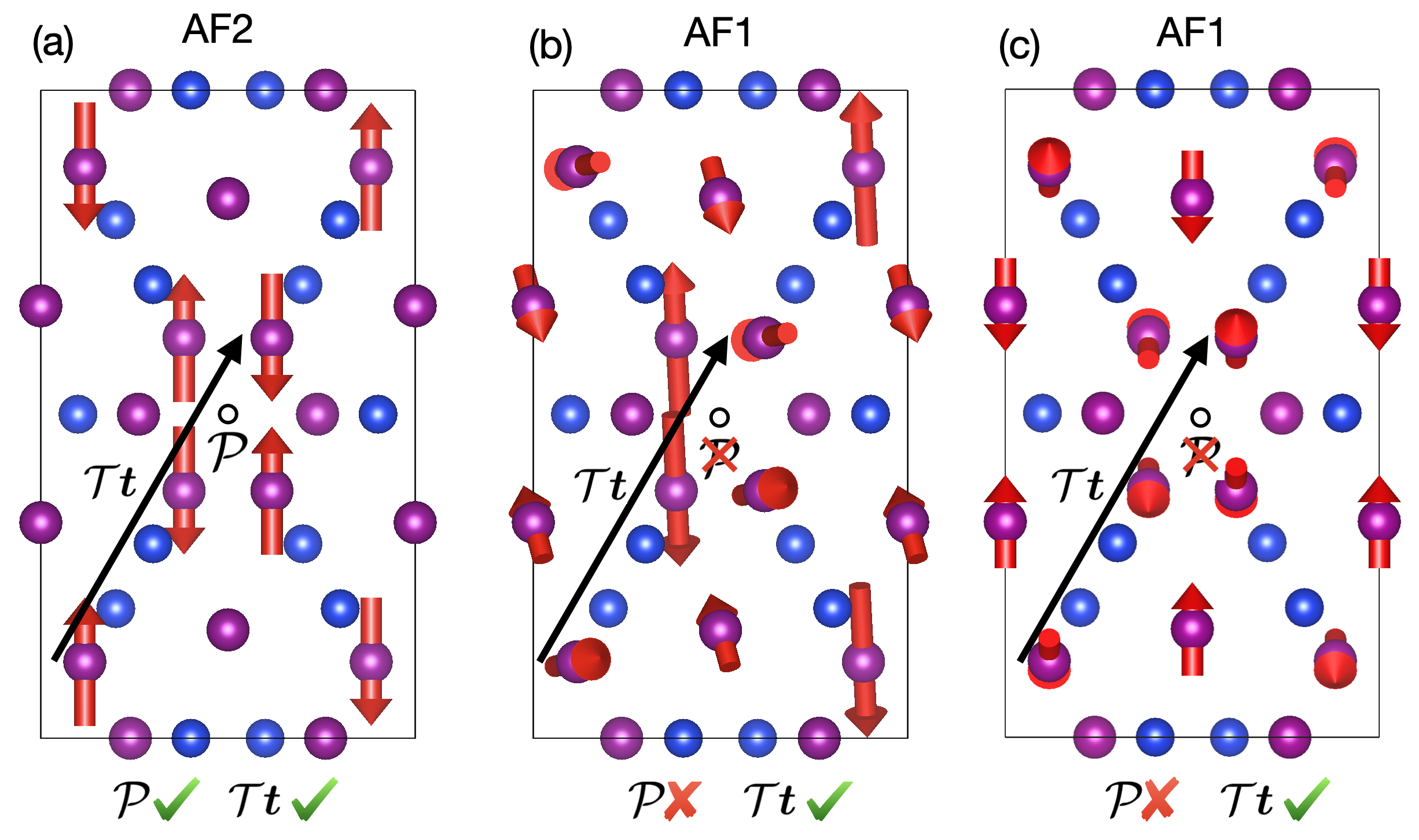}
	\caption{Magnetic crystal structures of the low-temperature ($T<100$~K) antiferromagnetic phases of bulk Mn$_5$Si$_3$. Mn and Si atoms are shown in magenta and blue color, respectively. (a) Collinear AF2 phase observed for 60~K~$<T<$100~K \cite{gottschilch_study_2012,brown_antiferromagnetism_1995,dos_santos_spin_2021,gallego2016}. (b) Noncoplanar AF1 phase without inversion symmetry proposed in Ref. \cite{brown_low-temperature_1992} for $T<60$~K \cite{gallego2016}. (c) Illustration of the AF1 phase proposed in Ref. \cite{biniskos_complex_2022} for $T<60$~K, also noncoplanar and noncentrosymmetric.}
	\label{fig:phases}
\end{figure}

Fig. \ref{fig1} shows the magnetic phase diagram of Mn$_5$Si$_3$ in a magnetic field $H$ oriented along the crystallographic \textit{c} axis. 
Various methods including neutron scattering have confirmed the existence of an antiferromagnetic phase AF2 between the N\'{e}el temperatures $T_{\rm N1}$ = 60 K and $T_{\rm N2}$ = 100 K with zero Mn$_1$ moments and collinear arrangement of two thirds of Mn$_2$ moments [Fig. \ref{fig:phases}(a)]  \cite{lander_antiferromagnetic_1967,brown_antiferromagnetism_1995,silva_magnetic_2002,gottschilch_study_2012,biniskos_spin_2018}. 
In the antiferromagnetic AF1 phase below $T_{\rm N1}$ = 60 K, a highly noncollinear and noncoplanar arrangement of magnetic moments is observed where the Mn atoms acquire different magnetic moments, although the details of the moment size and orientation of the Mn atoms are still under dispute \cite{brown_low-temperature_1992,bilbao_crystallographic_nodate,biniskos_complex_2022}, see Fig. \ref{fig:phases}(b,c). 
As soon as the magnetic structure changes from collinear AF2 to noncollinear AF1 below $T_{\rm N1}$ = 60 K, a strong AHE is observed \cite{surgers_large_2014,surgers_anomalous_2016,surgers_switching_2017}. 
In addition, the existence of a magnetic-field induced intermediate phase AF1´ has been inferred from neutron scattering and electronic transport measurements \cite{silva_magnetic_2002,surgers_switching_2017,biniskos_spin_2018}. 
Finally, at high magnetic fields, a collinear AF2 phase with similar properties like the zero field AF2 phase is reestablished \cite{surgers_large_2014,surgers_switching_2017,biniskos_spin_2018,biniskos_complex_2022}. 
The different phases show characteristic high-energy spin dynamics observed by inelastic neutron scattering \cite{biniskos_overview_2023}.
It is these different magnetic moment configurations of noncollinear and collinear order achieved at different magnetic fields and temperatures, that make the Mn$_5$Si$_3$ compound so interesting for investigations of the effect of topology and complex magnetic order on the magnetotransport properties. 

For the noncollinear AF1 phase, previous first-principle calculations assumed a Heisenberg Hamiltonian that takes into account magnetic exchange and biaxial anisotropy \cite{dos_santos_spin_2021}. 
Through modeling of the low-temperature spin-wave spectrum obtained by inelastinc neutron scattering at 10 K, two additional exchange constants representing the Mn$_1$-Mn$_1$ and Mn$_1$-Mn$_2$ interactions have been proposed resulting in a ground-state spin arrangement in the AF1 phase that is very different from the AF2 phase, with Mn$_1$ and Mn$_2$ moments oriented along the \textit{b} and \textit{c} axis, respectively [Fig. \ref{fig:phases}(c)] \cite{biniskos_complex_2022}. 
Furthermore, the field-induced phase transition between AF1 and AF1´ was simulated by including a Zeeman term in the Heisenberg Hamiltonian resulting in an additional magnetic phase to occur at low temperatures \cite{biniskos_complex_2022}. 
In this model, it is assumed that the Mn$_1$ moments are longitudinally susceptible, i.e., their size is affected by an external field. 
With increasing magnetic field a "spin flop" phase with coplanar moments in the \textit{ab} plane is succeeded by the AF1' phase, where Mn$_1$ moments start to acquire a component along the \textit{c} axis, and finally by a transition to the field-induced AF2 phase with nonvanishing Mn$_1$ moments aligned parallel to the direction of the magnetic field. 

In addition to single crystals, thin epitaxially strained Mn$_5$Si$_3$ layers with vanishing magnetization have been studied that exhibit collinear order and a spontaneous AHE due to the breaking of time-reversal symmetry by an unconventional staggered spin-momentum interaction \cite{reichlova_macroscopic_2021}. 
In this case, the zero net magnetization is generated by the non-relativistic electronic structure with altermagnetic collinear spin polarization in momentum space \cite{smejkal_giant_2022,smejkal_beyond_2022}.  
This gives rise to an anomalous Hall and Nernst effects despite the collinear spin arrangement in the thin epitaxial films \cite{badura_observation_2024}.

The link between the AHE and ANE via the Berry-phase concept, the hitherto reports of both effects observed in other noncollinear antiferromagnets, the observation of a nonzero AHE and its strong changes at the magnetic phase boundaries motivated this study of the ANE in Mn$_5$Si$_3$, where we focus to the noncollinear magnetic phase at low temperatures.

\section{Experimental}
\begin{figure}
	\includegraphics[width=0.9\columnwidth]{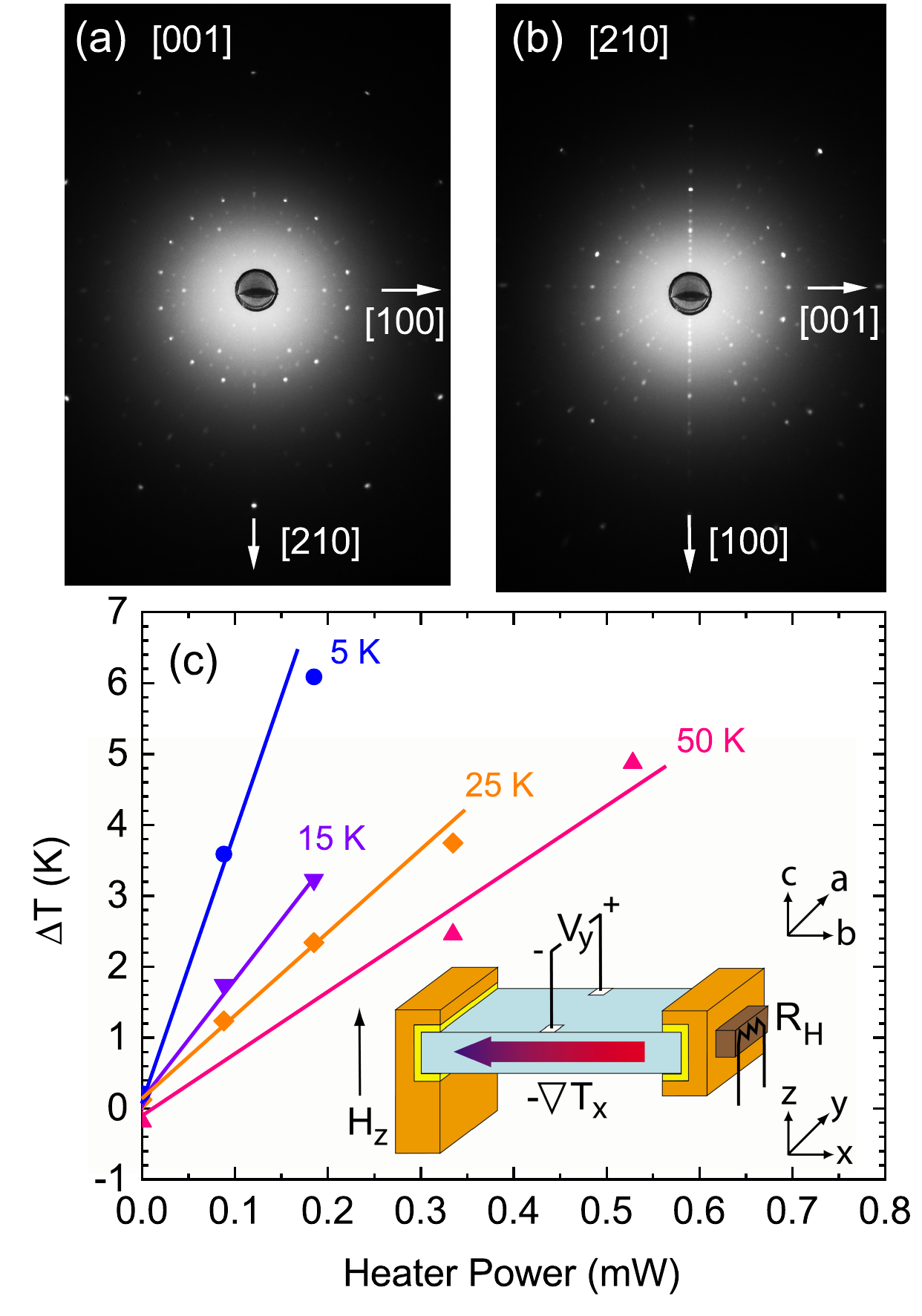}
	\caption[]{(a,b) Laue diffraction patterns of a Mn$_5$Si$_3$ single crystal oriented with the x-ray beam along the (a) [001] and (b) [210] directions of the hexagonal structure at room temperature. At $T <$\,100 K, the [100], [210], and [001] directions correspond to the $a$, $b$, and $c$ axes, respectively, of the orthorhombic phase. (c) Temperature difference $\Delta T = T_{\rm h}-T_{\rm c}$ between the hot and the cold end of the sample vs heater power for different average temperatures $T_0$. Solid lines indicate a linear behavior. Inset shows a cartoon of the experimental set up and the orientation of the sample with crystallographic axes \textit{a}, \textit{b}, and \textit{c} of low-temperature orthorhombic phase with respect to the sample holder.} 
	\label{fig3}
\end{figure}

Mn$_5$Si$_3$ single crystals were obtained by a combined Bridgman and flux-growth technique and were characterized by 
x-ray diffraction as described earlier \cite{surgers_switching_2017}. The single crystals have been polished and oriented by Laue diffraction, see Fig. \ref{fig3}(a,b).

Measurements of the AHE were performed in a physical-property measurement system (PPMS) as described in Ref. \cite{surgers_switching_2017} with 50-$\mu$m Pt wires attached to the crystal in an appropriate fashion with conductive silver-epoxy. The ANE was obtained on the same Mn$_5$Si$_3$ single crystal of thickness $t$ = 0.6 mm and lengths $l_{\rm x}$ = 1.2 mm and $l_{\rm y}$ = 1.4 mm along \textit{z}, \textit{x}, and \textit{y}, respectively. Fig. \ref{fig3}(c) shows a cartoon of the ANE setup. The crystal was mounted between two Cu clamps electrically isolated by 20-$\mu$m Kapton foil and a resistor $R_{\rm H}$ was attached on one clamp  serving as a heater generating a temperature difference $\Delta T_{\rm x} = T_{\rm h}-T_{\rm c}$ between the hot and cold side. Both temperatures were determined by using calibrated resistive thermometers. The power of the heater was always adjusted to keep $\Delta T_{\rm x}$ below 10\% of the average temperature $T_{\rm 0} = (T_{\rm h}+T_{\rm c})/2$. The Nernst voltage $\Delta V_{\rm y}$ was measured along the \textit{y} direction with a nanovoltmeter. The crystal was mounted with the crystallographic \textit{a}, \textit{b}, and \textit{c} axes oriented parallel to the \textit{y}, \textit{x}, and \textit{z} directions, respectively, with the heat flow $Q$ along the \textit{b} axis and the magnetic field oriented along the \textit{c} axis.  
 \begin{figure*}
	\includegraphics[width=2\columnwidth]{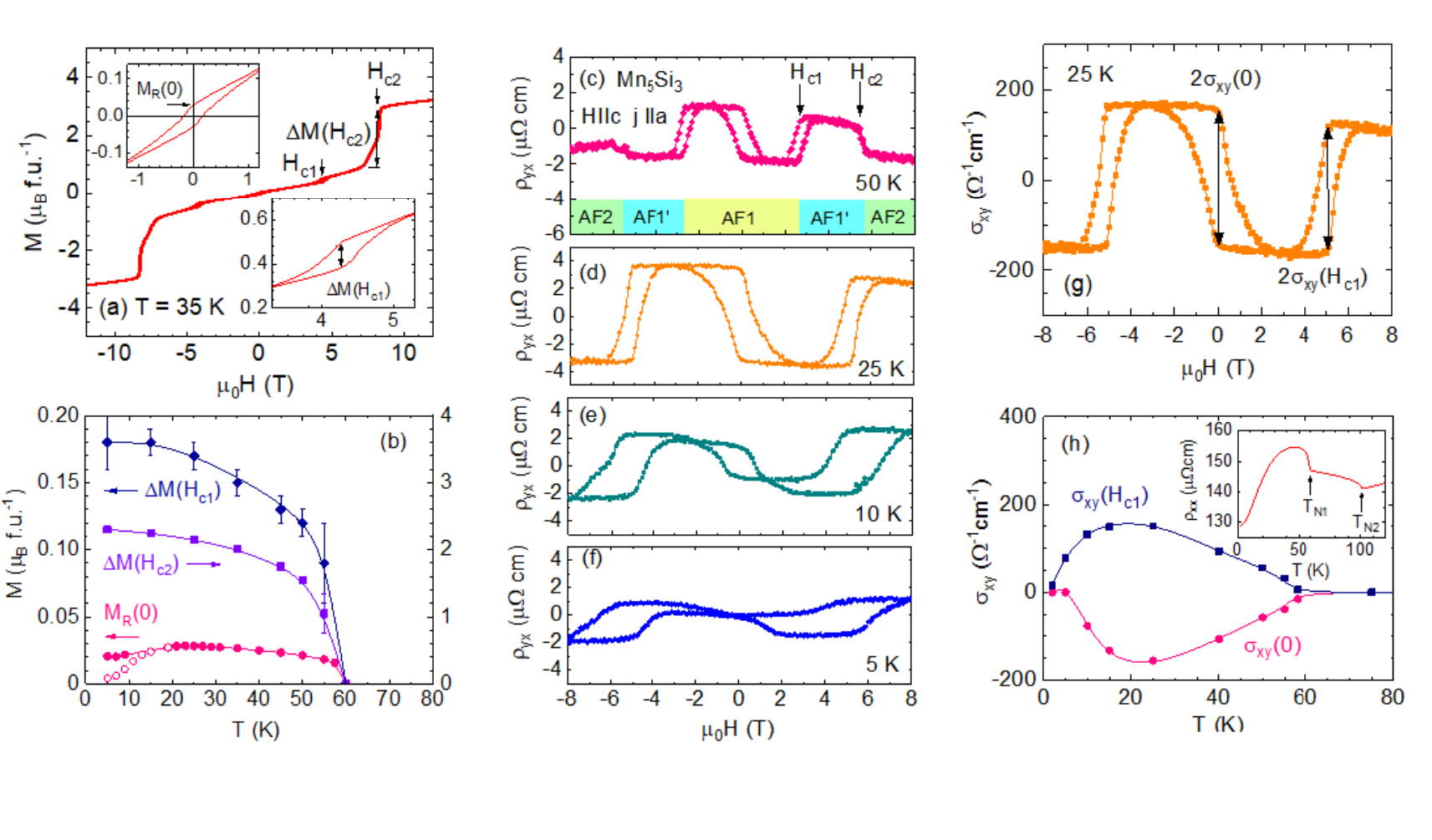}
	\caption[]{(a) Magnetization $M(H)$ at $T$ = 35 K. Insets show a close up of $M(H)$ near zero field (top left) and around the transition at $H_{\rm c1}$ (bottom right). (b) Remanent field $M_{\rm R}$(0) and magnetization jumps $\Delta M(H_{\rm c1})$ and $\Delta M(H_{\rm c2})$ obtained from $M(H$) vs temperature. Closed (open) circles indicate data obtained after applying a magnetic field of $\pm 5$ T ($\pm 2$ T) before measuring $M(0)$. (c-f) Anomalous Hall effect for various temperatures. (g) Anomalous Hall conductivity $\sigma_{\rm xy}(H)$ at $T$ = 25 K. (h) Temperature dependence of $\sigma_{\rm xy}(0)$ and $\sigma_{\rm xy}(H_{\rm c1})$. Inset shows the resistivity $\rho_{\rm xx}(T)$ with indicated N\'{e}el temperatures $T_{\rm N1}$ and $T_{\rm N2}$ at the magnetic phase transitions AF1/AF1’ and AF1’/AF2, respectively. The magnetic field was always oriented along the crystallographic \textit{c} axis.}
	\label{fig2}
\end{figure*}

\section{Results}
\subsection{Magnetization and Anomalous Hall effect}
Fig. \ref{fig2}(a) shows a typical magnetization curve of  Mn$_5$Si$_3$ at $T$ = 35 K with the two jumps of the magnetization $\Delta M(H_{\rm c1})$ and $\Delta M(H_{\rm c2})$ at magnetic fields $H_{\rm c1}$ and $H_{\rm c2}$, respectively, attributed to the AF1/AF1' and AF1'/AF2 phase phase transitions. 
The heights of the jumps are plotted in Fig. \ref{fig2}(b) together with the remanent magnetization $M_R(0)$ vs. temperature. 
 A remarkable detail that escaped our attention earlier is the decrease of $M_{\rm R}(0)$ when cooling below 25 K while $\Delta M(H_{\rm c1})$ and $\Delta M(H_{\rm c2})$ continuously increase with decreasing temperature below $T_{\rm N1}$ = 60 K. 
 Moreover, the absolute value of $M_{\rm R}(0)$ below 20 K depends on the maximum applied magnetic field, i.e. $\pm 2$ T or $\pm 5$ T, see Fig. \ref{fig2}(b), indicating the formation of antiferromagnetic domains with opposite N\'{e}el vectors in this temperature range \citealp{reichlova_macroscopic_2021}. Magnetic polarization at fields larger than 5 T was not possible without entering the AF1' phase.   
\begin{figure*}
	\includegraphics[width=2\columnwidth]{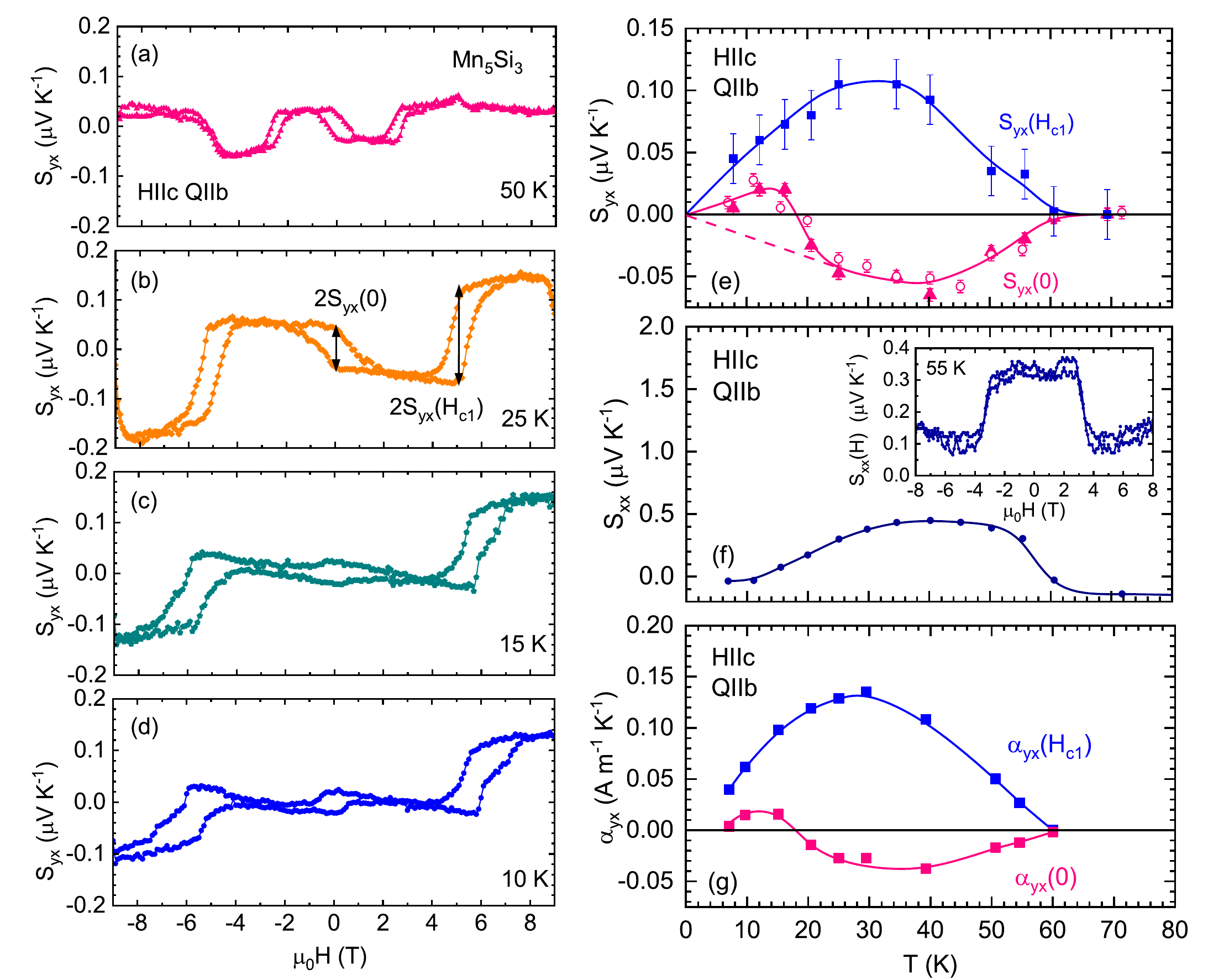}
	\caption[]{(a-d) Anomalous Nernst effect $S_{\rm yx}(H)$ for various temperatures. (e) $S_{\rm yx}(0)$ and $S_{\rm yx}(H_{\rm c1})$ vs temperature $T$. Solid symbols indicate data taken from individual $S_{\rm yx}(H)$ sweeps, see panels (a-d). Open circles were calculated from the difference between $S_{\rm yx}(H)$ measured in two consecutive temperature sweeps at positive or negative remanence. Dashed line indicates a linear behavior toward zero temperature. (f) Temperature dependence of the Seebeck coefficient $S_{\rm xx}$ in zero field. Inset shows $S_{\rm xx}(H)$ at $T$ = 55 K. (g) Transverse Peltier coefficients $\alpha_{\rm yx}(0)$ and $\alpha_{\rm yx}(H_{\rm c1})$ vs temperature $T$.}
	\label{fig4}
\end{figure*}  
  
The phase transitions are clearly observed in the AHE of Mn$_5$Si$_3$, see Fig. \ref{fig2}(c-f). 
As discussed earlier, the intrinsic Berry-phase contribution of the AHE to $\sigma_{\rm xy}$ is dominant in the range of moderate conductivity \cite{lee_hidden_2007,surgers_switching_2017}. 
At 50 K, below $T_{\rm N1}$, the AHE strongly changes at magnetic fields $H_{\rm c1}$ and $H_{\rm c2}$. 
In the collinear AF2 phase the AHE vanishes. 
The absolute value of the transverse resistivity $\rho_{\rm yx}(0)$ at zero field increases with decreasing temperature down to 25 K and then decreases and even vanishes at $T$ = 5 K while $M_{\rm R}(0)$ remains non zero. 

Fig. \ref{fig2}(g) shows the Hall conductivity $\sigma_{\rm xy} = \rho_{\rm yx}/(\rho_{\rm xx}^2+\rho_{\rm yx}^2)$ obtained by using the temperature dependence of the longitudinal resistivity $\rho_{\rm xx}$, see inset Fig. \ref{fig2}(h). 
The transverse conductivity $\sigma_{\rm xy} = 150\, \Omega^{-1}{\rm cm}^{-1}$ and the longitudinal conductivity $\sigma_{\rm xx} = 7400\, \Omega^{-1}{\rm cm}^{-1}$ are in agreement with the universal scaling relation of $\sigma_{\rm xy}$ and $\sigma_{\rm xx}$ \cite{chen_anomalous_2021,surgers_switching_2017,lee_hidden_2007}. 
  
The absolute values of $\sigma_{\rm xy}(0)$ and $\sigma_{\rm xy}(H_{\rm c1})$ plotted in Fig. \ref{fig2}(h) gradually increase between $T_{\rm N1}$ = 60 K and $\approx 25$\,K but then drop to very low values fro $T \le$\,25 K. 
We emphasize that $\sigma_{\rm xy}(0)$ completely vanishes for $T \le$ 5 K while $\sigma_{\rm xy}(H_{\rm c1})$ remains at a low value of 17 $\Omega^{-1}{\rm cm}^{-1}$ at $T =$ 2 K. 
The decrease of the AHE toward low temperatures is unusual for a well established magnetic order but could arise from domain reformation around zero field inferred from the reduced magnetization $M_R(0)$. 
However, the fact that both contributions to the AHE, $\sigma_{\rm xy}(0)$ and $\sigma_{\rm xy}(H_{\rm c1})$, decrease toward zero temperature while the magnetization $\Delta M(H_{\rm c1})$ is not reduced demonstrates that a different mechanism might be at hand.   
For antiferromagnetic Mn$_3$Sn, a sharp drop of $\sigma_{\rm zx}$ and $\sigma_{\rm yz}$ in connection with a strong increase of $\sigma_{\rm xy}$ was observed in the low temperature spin-glass phase below 50 K \cite{nakatsuji_large_2015,chen_anomalous_2021}. 
This suggests that in Mn$_5$Si$_3$ the magnetic moments possibly rearrange at low temperatures as previously inferred from analyzing the evolution of spin-wave energies, revealing another field induced phase-transition below the AF1/AF1' phase boundary \cite{biniskos_complex_2022}.

\subsection{Anomalous Nernst effect}
In the following we focus on the ANE obtained on the same Mn$_5$Si$_3$ single crystal. 
Theoretically, the ANE generates an electric field $\mathbf{E} = Q_{\rm S} \mu_0 \mathbf{M} \times (-\nabla T)$ perpendicular to the directions of the magnetization $\mathbf{M}$ and temperature gradient $-\nabla T$, where $Q_{\rm S}$ is the anomalous Nernst coefficient and $\mu_0$ is the magnetic permeability of free space. 
For the present configuration with $H_z$ parallel to the easy axis of magnetization \textit{c}, $\left| \mathbf{M}\right| = M_{\rm z}$, and with a temperature gradient along \textit{x}, this simplifies to $E_{\rm y} = -S_{\rm yx} \nabla T_{\rm x}$.  

The Nernst signal $S_{\rm yx}$ is experimentally measured by
\begin{equation}
S_{\rm yx} = \frac{\Delta V_{\rm y}\, l_{\rm x}}{\Delta T_{\rm x} \,l_{\rm y}}
\end{equation}
where we made use of the fact that $E_{\rm y} = -\Delta V_{\rm y}/l_{\rm y}$ and $\nabla T_{\rm x} = \Delta T_{\rm x} /l_{\rm x}$.

Fig. \ref{fig4}(a-d) shows $S_{\rm yx}$ vs magnetic field $H$ applied parallel to the \textit{c} direction and heat flow along the \textit{b} direction for different temperatures below $T_{\rm N1}$. 
At 50 K, the transitions between the different magnetic phases are observed as clear changes of $S_{\rm yx}$ with a hysteresis, very similar to the behavior of the AHE [Fig. \ref{fig2}(c-f)]. 
$S_{\rm yx}$ is zero in the collinear AF2 phase for temperatures $T > 60 $ K, see \textit{e.g.} Fig. \ref{appendix}(a), concomitant with a vanishing AHE and magnetization. 
Again, after attaining a maximum the magnitude of $S_{\rm yx}$ decreases with decreasing temperature below 25 K, similar to the behavior of the AHE, \textit{cf.} Fig. \ref{fig2}(c-f). 
However, a remarkable difference is the sign change of the remanent $S_{\rm yx}(0)$ below 25 K [see Figs. \ref{fig4}(b,c)] before it vanishes at lower temperatures. This behavior is also observed for $H$ applied parallel to the \textit{c} direction and heat flow along the \textit{a} direction but not for $H$ applied parallel to the \textit{a} or \textit{b} directions where no AHE appeared at zero field, see Appendix.  

The temperature dependence is seen more clearly in plots of $S_{\rm yx}(0,T)$ and $S_{\rm yx}(H_{\rm c1},T)$, Fig. \ref{fig4}(e). 
$S_{\rm yx}(0)$ rapidly changes sign by cooling to below 25 K, strongly deviating from the approximately linear temperature dependence observed for $S_{\rm yx}(H_{\rm c1})$ below 25 K.
Both coefficients vanish toward zero temperature due to Nernst´s theorem \cite{miyasato_crossover_2007}. 
The sign change of $S_{\rm yx}(0,T)$ occurs at the same temperature where we observe a decrease of the remanent magnetization $M_{\rm R}(0)$, in contrast to $\Delta M(H_{\rm c1})$ which appears more like a magnetically saturated state at low temperatures, see Fig. \ref{fig2}(b).
While the vanishing $\sigma_{\rm xy}(0)$ and reduced $M_{\rm R}(0)$ could be considered as indications for the reformation of 
antiferromagnetic domains, this cannot explain the reappearance of a positive $S_{\rm yx}(0)$ below 20 K supporting the idea of a weak modification of the magnetic structure at low temperatures.   

The transverse thermoconductivity, i.e. transverse Peltier coefficient, arising from the Berry phase at $E_{\rm F}$ \cite{xiao_berry-phase_2006}, can be calculated in zero magnetic field by
\begin{equation}
     \alpha_{\rm yx}= S_{\rm yx} \sigma_{\rm xx} - S_{\rm xx} \sigma_{\rm xy}
	\label{alpha}
\end{equation}
Here, $S_{\rm xx} = \Delta V_{\rm x}/\Delta T_{\rm x}$ is the Seebeck coefficient measured in the same setup and plotted in Fig. \ref{fig4}(f) for different temperatures using $\sigma_{\rm yx} = -\sigma_{\rm xy}$. 
An interesting detail is the magnetic field behavior of $S_{\rm xx}(H)$ in the inset of Fig. \ref{fig4}(f). At magnetic fields below $H_{\rm c2}$ the Seebeck coefficient is almost independent of $H$ but precipituously drops at $H_{\rm c2}$ when crossing the phase boundary between the field-induced noncollinear AF1' phase and the high-field collinear phase AF2. 
This extraordinary sharp change of the magneto-Seebeck effect with magnetic field is observed only at $H_{\rm c2}$ where the magnetic order changes from noncollinear to collinear. 
It is correlated with the strong change of the magnetoresistance $\rho_{\rm xx}(H)$ by $\approx$ 10 \% at $H_{\rm c2}$ and $T$ = 45 K \cite{surgers_switching_2017}. This is expected from the Mott relation between the thermo-electric and electronic conductivity
\begin{equation} 
	\alpha_{\rm xx} = \sigma_{\rm xx} S_{\rm xx} = \frac{\pi^2 k_{\rm B}^2}{3 e} T \left(\frac{\partial \sigma_{{\rm xx}}}{\partial E}\right)_{\mu}
	\label{Mott}
\end{equation}
where $k_{\rm B}$ is the Boltzmann constant, $e$ the electron charge, and $\mu$ the electrochemical potential. 	
A modification of the Fermi surface when crossing the AF1'/AF2 phase boundary can be also inferred from the strong variation of the carrier density obtained from the ordinary Hall effect of Mn$_5$Si$_3$ films \cite{surgers_large_2014}. 

It has been shown previously that the Mott relation between $\sigma_{\rm xx}$ and $\alpha_{\rm xx}$ can also be applied for the transverse coefficient $\alpha_{\rm xy} = -\alpha_{\rm yx}$ \cite{wang_onset_2001,xiao_berry-phase_2006,pu_mott_2008}
\begin{equation}
	\alpha_{\rm xy} = \frac{\pi^2 k_{\rm B}^2}{3 e} T \left(\frac{\partial \sigma_{{\rm xy}}}{\partial E}\right)_{\mu}.
	\label{Mottxy} 
\end{equation}

The transverse coefficient $\alpha_{\rm yx}$ [Fig. \ref{fig4}(g)] obtained from Eq. \ref{alpha} essentially shows the same behavior as $S_{\rm yx}(0)$ [Fig. \ref{fig4}(e)]. Since $\left| S_{\rm xx}\sigma_{\rm xy}\right| \ll \left|S_{\rm yx}\sigma_{\rm xx}\right|$, $\alpha_{\rm yx}$ is dominated by $S_{\rm yx} \sigma_{\rm xx}$ (Eq. \ref{alpha}). 
Hence, the ANE mostly arises from the transverse heat flow (90 \%) and the sign change of $\alpha_{\rm yx}$ occurs due to the sign change of $S_{\rm yx}$.

\section{Discussion}
\subsection{Temperature dependence of the ANE}
A sign change of the ANE with temperature was reported earlier for the ferromagnetic semiconductor Ga$_{1-x}{\rm Mn}_x$As 
and was attributed to the scattering-independent nature of the intrinsic AHE following a behavior $\rho_{\rm yx} \propto \rho_{\rm xx}^2$ \cite{pu_mott_2008}. 
The sign change of $S_{\rm yx}$ was found to be correlated with a sharp maximum in $S_{\rm xx}$ thus compensating this contribution to the transverse coefficient $\alpha_{\rm yx}$ which did not change sign. 
In contrast, for Mn$_5$Si$_3$ the sign change occurs for $S_{\rm yx}$ \textit{and} $\alpha_{\rm yx}$ but not for $S_{\rm xx}$. 
An analysis of the temperature dependence of $\alpha_{\rm xy}$ based on the power law $\rho_{\rm yx}^{\rm AHE} = \lambda M_z\rho_{\rm xx}^n$ \cite{pu_mott_2008} is not possible because $\lambda$ of Mn$_5$Si$_3$ is strongly temperature dependent \cite{surgers_large_2014} in contrast to itinerant ferromagnets where it is usually temperature independent at low temperatures.   

In the pyrochlore molybdate $R_2$Mo$_2$O$_7$ ($R$ = Nd, Sm) with noncoplanar spin structure the variations of $S_{\rm xy}$ and $\alpha_{\rm xy}$ were attributed to a contribution from the spin chirality of the system which was considered to be responsible for an enhanced AHE and a positive ANE at low temperatures \cite{hanasaki_anomalous_2008}. 
A strong magnetic field reduces the amplitude of the spin chirality by aligning the Mn moments along the magnetic field direction leading to a reduced $\sigma_{\rm xy}$ and $\alpha_{\rm xy}$.
It is interesting to note that this is observed only for $R$ = Nd, Sm with non-coplanar spin structure and not for collinearly 
ordered Gd$_2$Mo$_2$O$_7$.
In the present case, it is therefore conceivable that a similar behavior occurs due to a small change of the spin structure by moment reorientation and decrease of chirality/noncollinearity at small magnetic fields. 

We only mention that for the Weyl semimetal Co$_3$Sn$_2$S$_2$ different behaviors of $\alpha_{\rm yx}$ have been reported  \cite{ding_intrinsic_2019,guin_zero-field_2019,yang_giant_2020}. 
In these cases, however, $S_{\rm xy}$ did not change its sign with the temperature.  

For Mn$_5$Si$_3$, maximum values $\left|S_{\rm yx}(0) \right|$ = 0.05 $\mu {\rm VK}^{-1}$ and $\left|S_{\rm yx}(H_{\rm c1}) \right|$ = 0.11 $\mu {\rm VK}^{-1}$ are reached at magnetizations $M_{\rm R}(0) = 0.028\, \mu_{\rm B}$/f.u. and $M_{\rm R}(H_{\rm c1}) = 0.17\, \mu_{\rm B}$/f.u., respectively. 
Note that the former is a lower limit due to the thermal resistances between the sample holder and the single crystal which give rise to a lower applied thermal gradient in the crystal than the measured $\Delta T_{\rm x} = T_{\rm h} - T_{\rm c}$. 
Hence, an even larger $\left|S_{yx} \right|$ derived from the low magnetization is expected. 
Similar values $\left|S_{yx} \right| = 0.6 \, \mu{\rm VK}^{-1}$ and $\mu_0M$ = 1 mT corresponding to $M = 0.01\, \mu_{\rm B}$/f.u. have been reported for chiral Mn$_3$Sn \cite{ikhlas_large_2017}. 
Compared to ferromagnetic metals, $\left|S_{yx} \right|$ is strongly enhanced outside the broad range for which $\left|S_{yx} \right| \propto M$ is observed, similar to antiferromagnets with chiral magnetic order like Mn$_3$Sn \cite{ikhlas_large_2017,chen_anomalous_2021} and Mn$_3$Ge \cite{wuttke_berry_2019,chen_anomalous_2021}.

Below $T_{\rm N1}$ = 60 K, $\alpha_{yx}(H_{\rm c1})$ first increases with decreasing temperature together with a concomitant increase of $\Delta M(H_{\rm c1})$, eventually saturates and then decreases towards low temperatures. 
This is similar to the behavior of an itinerant ferromagnet and due to the dominant $T$-linear term in Eq. \ref{Mottxy} after saturation of $M$ \cite{miyasato_crossover_2007}. 
In contrast, in the AF1 phase below $H_{\rm c1}$ the ANE strongly deviates from this behavior at temperatures below $\approx$\,25 K.  
We speculate that the reduction of $M_{\rm R}(0)$ and the simultaneous sign changes of $\alpha_{yx}(H_{\rm c1})$ and $S_{yx}(0)$ with temperature could arise from the different compensation of opposed Mn moments. 
In this respect it bears some similarity to the magnetization behavior of garnets or rare-earth transition-metal ferrimagnets below and above the compensation point. 

\subsection{Symmetry analysis}
To gain further insights into the low-temperature behavior of Mn$_5$Si$_3$, we analyze the symmetries of the anomalous Hall and Nernst effects under a small external magnetic field. 
The Onsager relations establish strong symmetry requirements for each response tensor. 
The (thermo-)electric responses measured in our experiments are approximately odd under the reversion of a small external magnetic field (Figs. \ref{fig2} and \ref{fig4}). 
Up to second order, the only odd-in-field transverse electric (thermo-electric) responses are the intrinsic and quadratic anomalous Hall (Nernst) effects \cite{grimmer_thermoelectric_2017}. 
These response tensors are even under space inversion $\mathcal{P}$ and odd under the combined time reversal and lattice translation symmetry $\mathcal{T}\bm{t}$ \cite{grimmer_thermoelectric_2017,Smejkal2020}. 
According to Neumann's principle, the transport coefficients must also be invariant under all the symmetries of the material. 
In the next paragraph, we combine the implications of the Onsager relations with Neumann’s principle to investigate the symmetry constraints of the AHE and ANE in the antiferromagnetic phases  AF1 and AF2 of Mn$_5$Si$_3$ (Fig. \ref{fig:phases}).

Consider the collinear AF2 phase shown in Fig. \ref{fig:phases}(a). 
The system is invariant under both $\mathcal{P}$ and $\mathcal{T}\bm{t}$. 
Because the response tensors are also invariant under $\mathcal{P}$, this symmetry operation does not impose any constraint. 
On the other hand, the $\mathcal{T}\bm{t}$ invariance of the system imposes that the response coefficients be even under time reversal. 
Since the Onsager relations require the response tensors to be odd under $\mathcal{T}$, the intrinsic and quadratic anomalous Hall/Nernst effects are prohibited in the AF2 phase, in agreement with our experimental results for $60$~K~$<T<100$~K [Figs. \ref{fig2}(h) and \ref{fig4}(e,g)] \cite{surgers_large_2014,surgers_anomalous_2016,surgers_switching_2017}. 
As mentioned above, the  magnetic configurations of  Mn$_5$Si$_3$ for $T<60$~K are still under discussion \cite{biniskos_complex_2022}.
The AF1 phases shown in Fig. \ref{fig:phases}(b) and Fig. \ref{fig:phases}(c) break $\mathcal{P}$ but (as well as other proposals in the literature) are invariant under $\mathcal{T}\bm{t}$, which allows for a p-wave non-relativistic spin splitting of the energy bands \cite{Hellenes2024}. 
However, even though the bands are not spin-degenerate, the $\mathcal{T}\bm{t}$ symmetry prohibits the anomalous responses. 
We conclude that, due to $\mathcal{T}\bm{t}$ invariance, none of the proposed AF1 phases can explain our experimental signals. 
This suggests a slight tilting of the magnetic state for $T<60$~K, which breaks the combined $\mathcal{T}\bm{t}$ symmetry. 
Below $T<25$~K, the reduction in the magnitude of the anomalous Hall conductivity and sign change of the anomalous Nernst conductivity hints to either a new phase or a weak rearrangement of the spins due to magnetic frustration or magnetic anisotropies, as pointed out by N. Biniskos \emph{et al.} \cite{biniskos_complex_2022}.

\section{Conclusion}
The anomalous Nernst effect of Mn$_5$Si$_3$ single crystals displays characteristic variations with magnetic field and temperature in agreement with the magnetic-field induced transitions between different antiferromagnetic phases. 
Detailed analysis of the low-temperature behavior below 25 K shows that the anomalous Nernst effect, anomalous Hall conductivity, and magnetization in low magnetic fields exhibit an unusual temperature dependence hinting a subtle modification of the magnetic structure in the AF1 phase.
Furthermore, the experimentally observed Hall and Nernst effects in the noncollinear AF1 phase are in contrast to a symmetry analysis of the proposed magnetic AF1 structures, according to which these effects should disappear.     
These results should be taken into account in a refined model of the magnetic structure.
While first ab initio calculations of the electronic band structure and Berry curvature have been performed for the collinear magnetic phase observed in thin films \cite{reichlova_macroscopic_2021,smejkal_giant_2022}, the behavior of the AHE and ANE in the noncollinear AF1 phase of Mn$_5$Si$_3$ at low temperatures is not yet fully understood and requires further investigation.
\vspace{8mm}
 
\section*{Acknowledgements}
This work was supported by the Deutsche Forschungsgemeinschaft (DFG) through CRC TRR 288 - 422213477 “ElastoQMat” (Projects A08, A09). 

\bibliography{Mn5Si3_ANE}

\appendix
\section{Supplementary data}
\renewcommand\thefigure{\thesection.\arabic{figure}}    
\setcounter{figure}{0}  
\begin{figure*}
	\includegraphics[width=2\columnwidth]{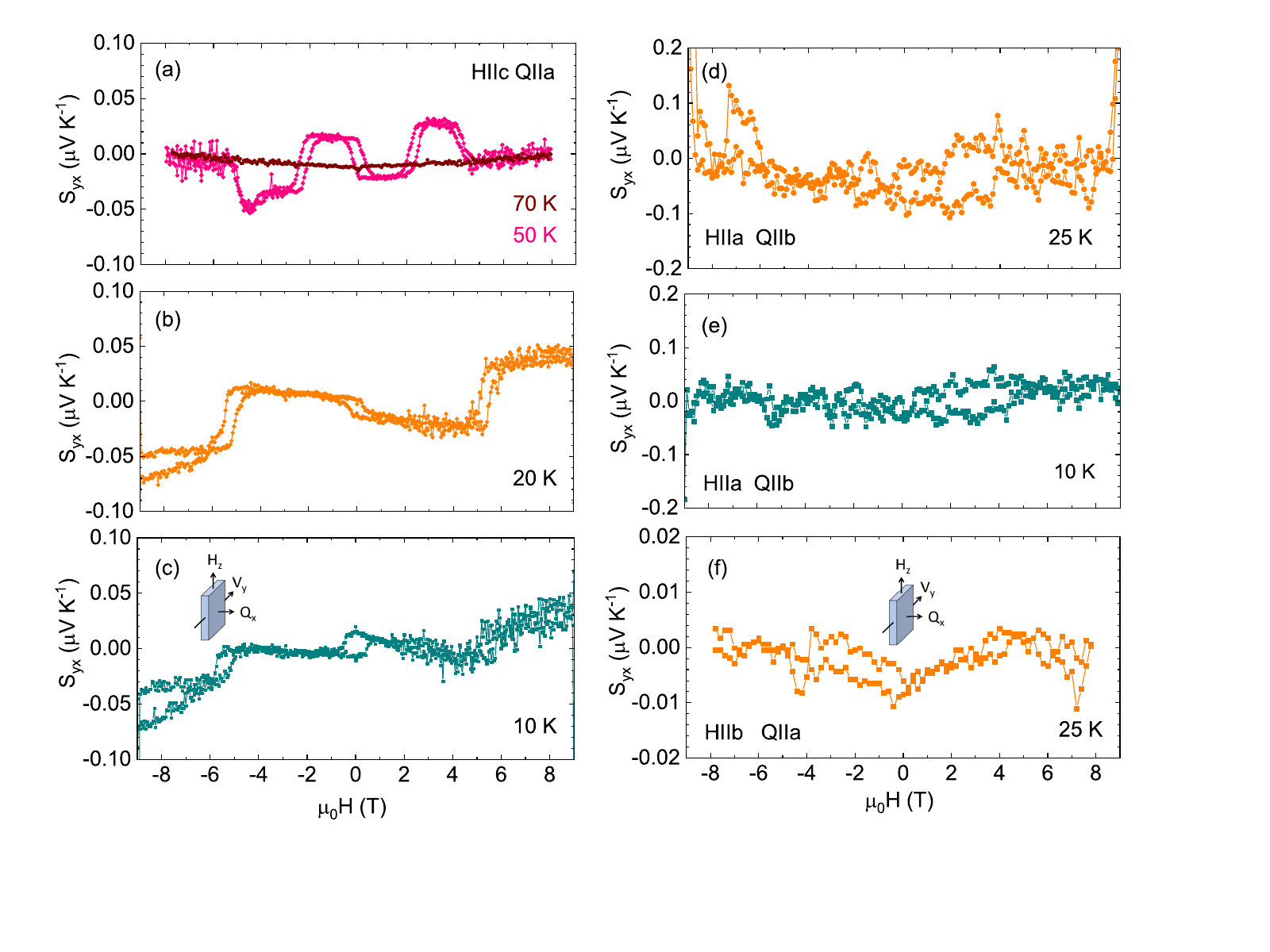}
	\caption[]{Anomalous Nernst effect of Mn$_5$Si$_3$ for magnetic fields \textit{H} and heat flows \textit{Q} applied along different crystallographic directions. (a-c) \textit{H} along \textit{c}, heat flow along \textit{a}. (d-f) Measurements with \textit{H} applied along the orthorhombic \textit{a} or \textit{b} direction. The measurements with heat flow along the \textit{a} direction were obtained on a thin plate of 0.52 mm thickness along x.}
	\label{appendix}
\end{figure*} 

\end{document}